\documentclass[aps,prb,10pt,twocolumn]{revtex4-1}
\usepackage{bm}
\usepackage{amsmath}
\usepackage{amsfonts}
\usepackage{amssymb}
\usepackage{graphicx}
\usepackage[colorlinks=true,linktocpage=true,breaklinks=true]{hyperref}

\begin{document}

\title{Non-equilibrium spin density and spin-orbit torque in three dimensional topological insulators - antiferromagnet heterostructure}
\author{S. Ghosh}
\email{sumit.ghosh@kaust.edu.sa}
\author{A. Manchon}
\email{aurelien.manchon@kaust.edu.sa}
\affiliation{King Abdullah University of Science and Technology (KAUST), Physical Science and Engineering Division (PSE), Thuwal 23955, Saudi Arabia}

\begin{abstract}
We study the behavior of non-equilibrium spin density and spin-orbit torque in a topological insulator - antiferromagnet heterostructure. Unlike ferromagnetic heterostructures where Dirac cone is gapped due to time-reversal symmetry breaking, here the Dirac cone is preserved. We demonstrate the existence of a staggered spin density corresponding to a damping like torque, which is quite robust against the scalar impurity, when the transport energy is in the topological insulator surface energy regime. We show the contribution to the non-equilibrium spin density due to both surface and bulk topological insulator bands. Finally, we show that the torques  in topological insulator-antiferromagnet heterostructure exhibit an angular dependence that is consistent with the standard spin-orbit torque obtained in Rashba system with some additional nonlinear effects arising from the interfacial coupling.
\end{abstract}

\maketitle

\section{Introduction}
Spin-orbit torque (SOT) has recently become a viable candidate mechanism for the development of magnetic memory devices \cite{Kent2015, Lee2016, Sato2018}.  A typical SOT device mainly consists of two elements, a source of strong spin-orbit coupling (SOC) and a magnetic material \cite{Miron2011, Liu2012}. When a charge current is passed through the material with strong SOC it produces a non-equilibrium spin density which is utilized to manipulate the magnetic order \cite{Manchon2018}. In recent years, the field of SOT devices has been revolutionized by two major breakthroughs: one is the discovery of topological insulator  (TI) \cite{Hasan2010a, Qi2011} and the other is the introduction of antiferromagnet (AF) \cite{Nunez2006, Gomonay2014, Jungwirth2016, Baltz2018}. The strong interfacial SOC in TIs provides a high charge-spin conversion efficiency and thus can be used as an efficient source of SOTs \cite{Mellnik2014, Fan2014, Wang2015, Fan2016a}. Recent experiments show that TI-based SOT devices can be operated with current of the order of $10^5 \rm A/cm^2$ even in room temperature \cite{Han2017, Wang2017, DC2018}, which is two orders of magnitude smaller than the switching current density required to operate heavy metal-based SOT devices.
Antiferromagnetic materials, on the other hand, have been known for decades and used extensively as passive exchange bias layers in spintronics spin-valves \cite{Nogues1999}. Ten years ago, it was proposed that the antiferromagnetic order parameter could be manipulated electrically using spin transfer torque \cite{Nunez2006, Gomonay2010}. The recent prediction that SOT could be used to control collinear antiferromagnet \cite{Zelezny2014, Zelezny2017, Manchon2017, Watanabe2018} confirmed shortly after in CuMnAs, opened appealing avenues as antiferromagnets are immune to external magnetic fields and host ultrafast (THz) dynamics \cite{Cheng2015, Cheng2016}. To date, SOT-driven switching has been observed in the noncentrosymmetric $\rm CuMnAs$ \cite{Wadley2016} and $\rm Mn_2Au$ crystals \cite{Bodnar2018, Meinert2018}, but also in $\rm Pt/NiO$ \cite{Chen2018} and compensated $\rm Pt/CoGd$ bilayers \cite{Moriyama2018}.
 The switching is not only robust against external magnetic field but takes place at much faster rate compared to the ferromagnet based SOT devices \cite{Roy2016, Olejnik2018, Bhattacharjee2018}. 

Due to their individual strength, these two fields of research have stimulated substantial amount of theoretical and experimental research. However there are very little attempt to combine these two fields. 
Considering the promises born by these two classes of materials, it is natural to investigate the nature of SOTs in AF-TI bilayers and identify its main features.
 One major difficulty in this regard is to determine the right material  combination, as interfacing TIs with transition metal layers is known to significantly affect the topological surface states (e.g., \onlinecite{Zhang2016, Tejada2017}).
Nonetheless, recent experimental progress has been achieved towards the fabrication of AF-TI heterostructures, revealing complex magnetic configuration at the interface and suggesting viable routes towards the observation of SOTs \cite{He2017, He2018}. A proper theoretical understanding of these systems is therefore highly solicited at this moment.      

In this work, we present a systematic study of spin transport in AF-TI heterostructures based on a tight-binding model. We calculate the non-equilibrium spin density using linear response theory and investigate its behavior with respect to the different system's parameters.
We show the existence of a non-equilibrium staggered spin density, localized near the interface and immune to scalar impurity at low energy. Similar behavior is observed for the longitudinal conductivity at energy close to the Dirac cone of the TI. The non-equilibrium spin densities show an angular dependence similar to that of a Rashba ferromagnet which can be distorted by the interfacial AF-TI coupling. 

\section{Method}
\begin{figure}[h]
\centering
\includegraphics[width=0.48\textwidth]{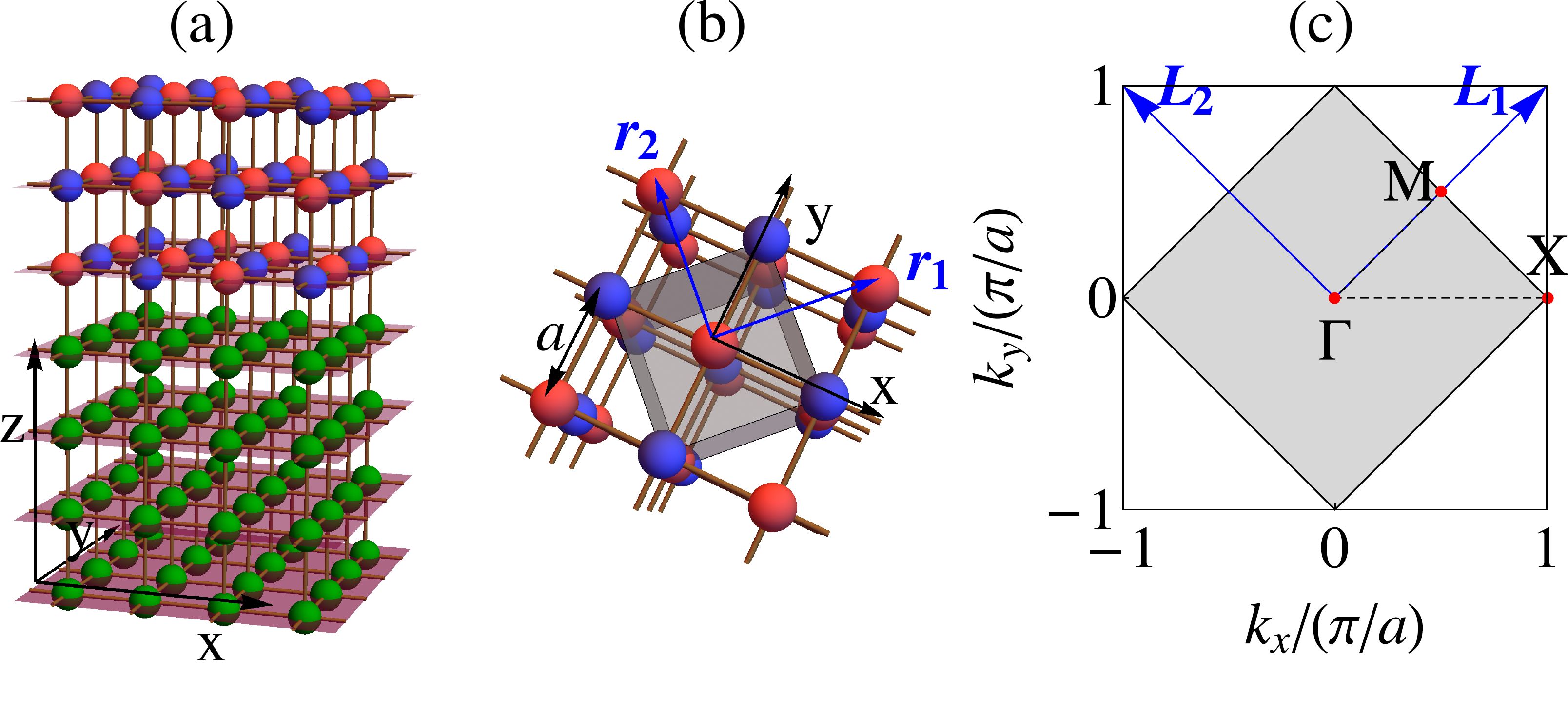}
\caption{(a) Schematic of an AF-TI heterostructure. Red and blue spheres denotes up and down magnetic moment of the AF layer and the green spheres denotes the TI. (b) Top view of an AF lattice. The gray boundary denotes the unit cell. $a$ is the interatomic distance. (c) Corresponding first Brillouin zone (gray region) with lattice vector $\bm L_1$ and $\bm L_2$ and high-symmetric points $M$, $\Gamma$ and $X$ for the AF-TI heterostructure.}
\label{fig1}
\end{figure}

We use a tight binding model to describe the AF-TI heterostructure. The TI is modeled following the   method described in Ref. \onlinecite{Ghosh2018} with the same set of parameters. To define the G-type antiferromagnet we have to double the unit cell (Fig.~\ref{fig1}b) and define lattice vectors $\bm{r_1}=(\bm{e_x}+\bm{e_y})$ and $\bm{r_2}=(\bm{e_x}-\bm{e_y})$ where $\bm e_x$ and $\bm e_y$ are the unit vector along the $x$ and $y$ axis. The corresponding first Brillouin zone and reciprocal lattice vectors ($\bm L_1$, $\bm L_2$) are shown in Fig.~\ref{fig1}c.

\begin{figure}[h]
\centering
\includegraphics[width=0.45\textwidth]{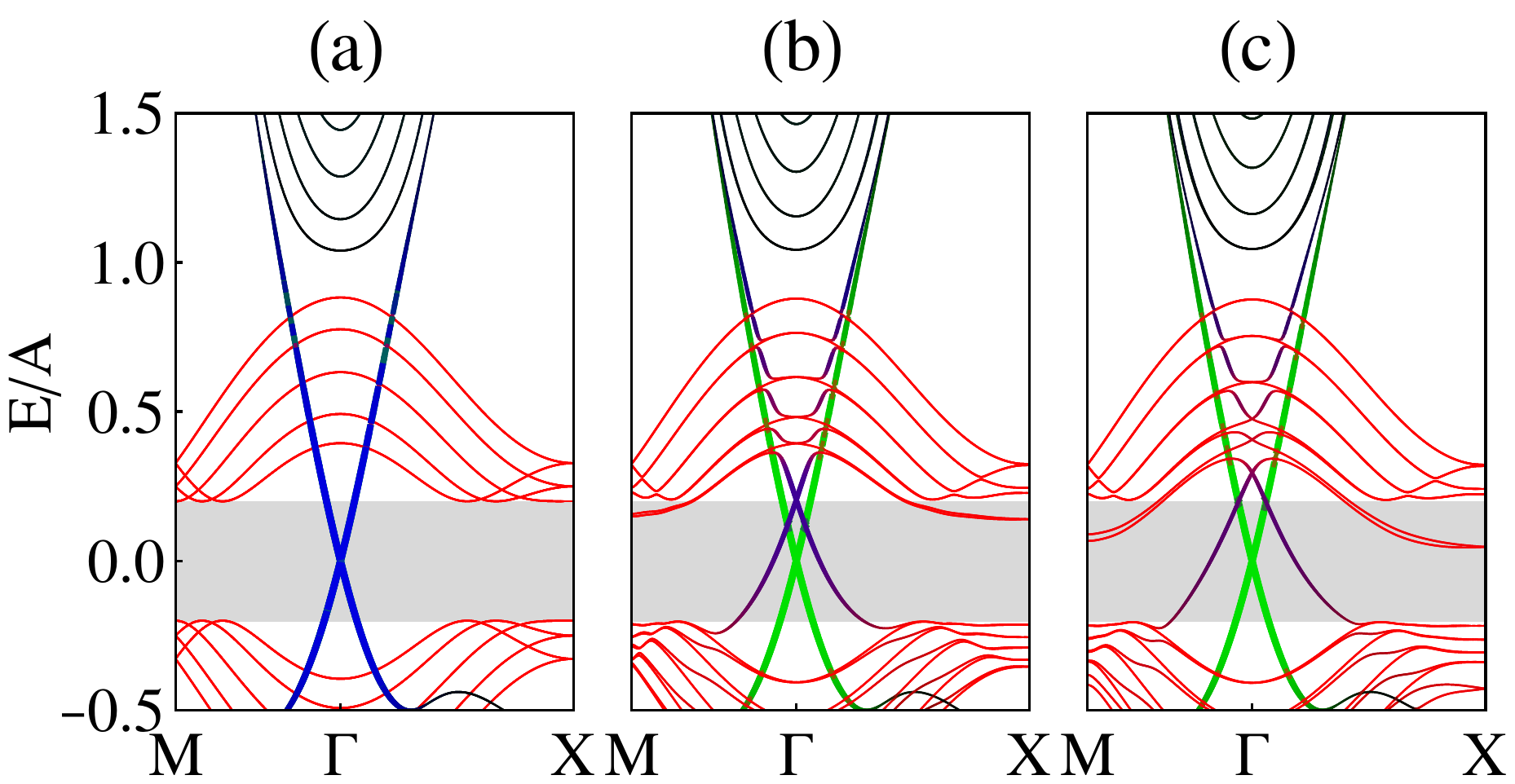}
\caption{Band structure of a heterostructure with 10 layers of TI and 5 layers of AF with a coupling strength (a) $t_C=0$, (b) $t_C=0.5$ and (c) $t_C=0.75$. The red, green, blue and black colors correspond to the contributions from AF layers, bottom TI layer, interfacial TI layer and bulk TI layers, respectively. The gray region shows the gap due to exchange splitting for decoupled AF.}
\label{fig:band}
\end{figure}

The Hamiltonian for the TI layers can be written as
\begin{eqnarray}
H_{TI} = \sum_{i} a^\dagger_i(h_0)a_i + \sum_{\langle i,j \rangle} a^\dagger_i (h_{ij}) a_j ,
\end{eqnarray}
where $a^\dagger_i/a_i$ is the creation/annihilation operator for the TI at $i^{th}$ site, $\langle \rangle$ denotes the summation over the nearest neighbors and $h_0$ and $h_{ij}$ denote the onsite and hopping elements. For a TI the onsite and hopping  elements are $4\times4$ matrices defined as
\begin{eqnarray}
 h_0 &=&  \Gamma_1 M + \mathbb{I}_4 c, \nonumber \\
 h_{-x}^\dagger =  h_{x} &=& (-i  \Gamma_2 A -  \Gamma_1 B - \mathbb{I}_4 d)/2, \nonumber \\
h_{-y}^\dagger = h_{y} &=& (i  \Gamma_3 A -  \Gamma_1 B - \mathbb{I}_4 d)/2, \nonumber \\
 h_{-z}^\dagger =  h_{z} &=& (-i  \Gamma_4 A_1 -  \Gamma_1 B_1)/2,
\end{eqnarray}   
where $\mathbb{I}_4$ is the identity matrix of rank 4, while $M$, $A$, $B$, $c$, $d$, $A_1$ and $B_1$ are the material parameters \cite{Ghosh2018}. The $\Gamma$ matrices are defined as
\begin{eqnarray}
&\Gamma_1 = \sigma_1 \otimes \mathbb{I}_2, \ & \Gamma_2 = \sigma_3 \otimes \sigma_2, \nonumber \\
& \Gamma_3 = \sigma_3 \otimes \sigma_1, \  & \Gamma_4 = \sigma_2 \otimes \mathbb{I}_2,
\end{eqnarray}
where $\mathbb{I}_2$ is the identity matrix of rank 2 and $\sigma$'s are the Pauli matrices for spin 1/2. 
Similarly the Hamiltonian for the AF layers is given by
\begin{eqnarray}
H_{AF} = \sum_{i} c^\dagger_i(\Delta_i \bm{m}.\bm{\sigma} + \varepsilon_0 \mathbb{I}_2) c_i - \sum_{\langle i,j \rangle} c^\dagger_i (t_{ij} \mathbb{I}_2) c_j,
\label{H_AF}
\end{eqnarray} 
where $c^\dagger_i/c_i$ is the creation/annihilation operator for the AF at $i^{th}$ site, $\Delta_i = \Delta(-1)^{x_i+y_i+z_i}$ is the staggered onsite exchange energy, and $\varepsilon_0$ is the uniform onsite energy which can tune the position of the AF bands. In case of AF we use  isotropic real hopping as $t_{x,y,z} = t_{-x,-y,-z} = t_{AF}$. The TI and AF layers are connected by 
\begin{eqnarray}
H_{AFTI} &=& \sum_{\langle i,j \rangle} a^\dagger_i (T_C) c_j + H.C. , \nonumber \\
T_C &=& t_C(\mathbb{I}_2 , \mathbb{I}_2),
\end{eqnarray}
where $H.C.$ denotes the Hermitian conjugate.
The typical parameter values we use here are $A=1.0$, $B=1.5$, $M=3.5$, $c=1.5$, $d=0.75$, $A_1=1.5$, $B_1=1.5$, $t_{AF}=0.15$ and $\Delta=0.2$. Unless explicitly mentioned otherwise, we use $\varepsilon=0.0$ and $t_C=0.5$ and set the magnetic moment of the AF layer out of plane. We consider here a system with 10 layers (20 sites) of TI and 5 layers (10 sites) of AF. Corresponding band structures for different coupling strengths are shown in Fig.~\ref{fig:band}. Note that the Dirac cone coming from the interfacial TI layer is still preserved \cite{Mong2010} and, depending on the AF-TI coupling strength, is shifted to slightly higher energy. In addition, we notice that the AF gap reduces upon increasing the coupling with the TI surface states.

The nonequilibrium spin density and conductivity are calculated within the linear response framework \cite{Freimuth2014, Li2015, Wimmer2016, Ghosh2018}. We start by defining the retarded/advanced Green's function $G^{R,A}(E,{\bm{k}})$ at energy $E$ and momentum $\bm{k}$,
\begin{eqnarray}
G^{R,A}(E,{\bm{k}})=[(E \pm i\eta){\mathbb{I}}_n - H({\bm{k}})]^{-1}.
\end{eqnarray}
For simplicity we write it as $G^{R,A}$ in the following section omitting the explicit dependence of $E$ and $\bm k$. The expectation value of an observable $\mathcal{O}$ due to a perturbation $\mathcal{P}$ consists of two parts  
\begin{eqnarray}
\langle \hat{\mathcal{O}}^{\hat{\mathcal{P}}} \rangle = \langle \hat{\mathcal{O}}^{\hat{\mathcal{P}}} \rangle_{\rm sur} + \langle \hat{\mathcal{O}}^{\hat{\mathcal{P}}} \rangle_{\rm sea},
\label{s}
\end{eqnarray}
where $\mathcal{O}_{\rm sur}$ and $\mathcal{O}_{\rm sea}$ correspond to contributions from Fermi surface and Fermi sea, respectively, and are defined by,
\begin{eqnarray}
\langle \hat{\mathcal{O}}^{\hat{\mathcal{P}}} \rangle_{\rm sur} &=& \frac{1}{2 \pi} \int \frac{d^2 k}{(2\pi)^2} ~ {\rm Re}({\rm Tr} [ \hat{\mathcal{O}} G^R \hat{\mathcal{P}} (G^A-G^R)])_{E_F}, \label{w1} \nonumber \\
\\
\langle \hat{\mathcal{O}}^{\hat{\mathcal{P}}} \rangle_{\rm sea} &=& \frac{1}{2 \pi} \int_{-\infty}^{E_F} dE \int \frac{d^2 k}{(2\pi)^2} ~ {\rm Re}\left( {\rm Tr} [ \hat{\mathcal{O}} G^R \hat{\mathcal{P}} \frac{\partial G^R}{\partial E} \right. \nonumber \\
&&\hspace{3.5cm} \left. - \hat{\mathcal{O}} \frac{\partial G^R}{\partial E} \hat{\mathcal{P}} G^R] \right) \label{w2}.
\end{eqnarray} 

In this article we are interested in the non-equilibrium spin density and conductivity due to an applied electric field applied along $x$ direction. Therefore the perturbation term is given by $\hat{\mathcal{P}} = e\mathcal{E}v_x$, where $v_x = \partial H /\partial (\hbar k_x)$ and $\mathcal{E}$ is the amplitude of the applied electric field and $e$ is the electrical charge. Since we are interested in the response function, the computed quantities are normalized with respect to E. The spin operator for individual site is given by
\begin{eqnarray}
\hat{\bm{S}}_i = \bm{s} \otimes |i \rangle \langle i|,
\end{eqnarray}   
where $\bm{s}=\bm{\sigma}$ for the AF and $\bm{s}=\bm{\sigma} \otimes \mathbb{I}_2$ for TI sites. $|i \rangle \langle i|$ is the projection operator for $i^{th}$ site. Since we are using a bipartite lattice, instead of site resolved velocity operator, we use an average velocity operator defined as
\begin{eqnarray}
\hat{v}_x^j = (1/\hbar[\partial H/ \partial k_x]_{...[layer_j]...})/2,
\end{eqnarray}  
where $...[layer_j]...$ refers to the fact that we take the block corresponding to the layer containing $j^{th}$ site from the full matrix. Since each layer contains two sites, we divide this matrix by 2 to obtain the average contribution coming from a single site. The response function for non-equilibrium spin density and the conductivity are given by
\begin{eqnarray}
\bm{S}_i &=& \langle \hat{\bm{S}}_i^{e\hat{v_x}} \rangle, \\
\sigma_{xx}^{j} &=& \langle e (\hat{v}^j_x)^{e\hat{v}_x} \rangle .
\end{eqnarray}

\section{Result and discussion}

\subsection{Spatial distribution of spin density and conductivity}

\begin{figure}[h]
\centering
\includegraphics[width=0.48\textwidth]{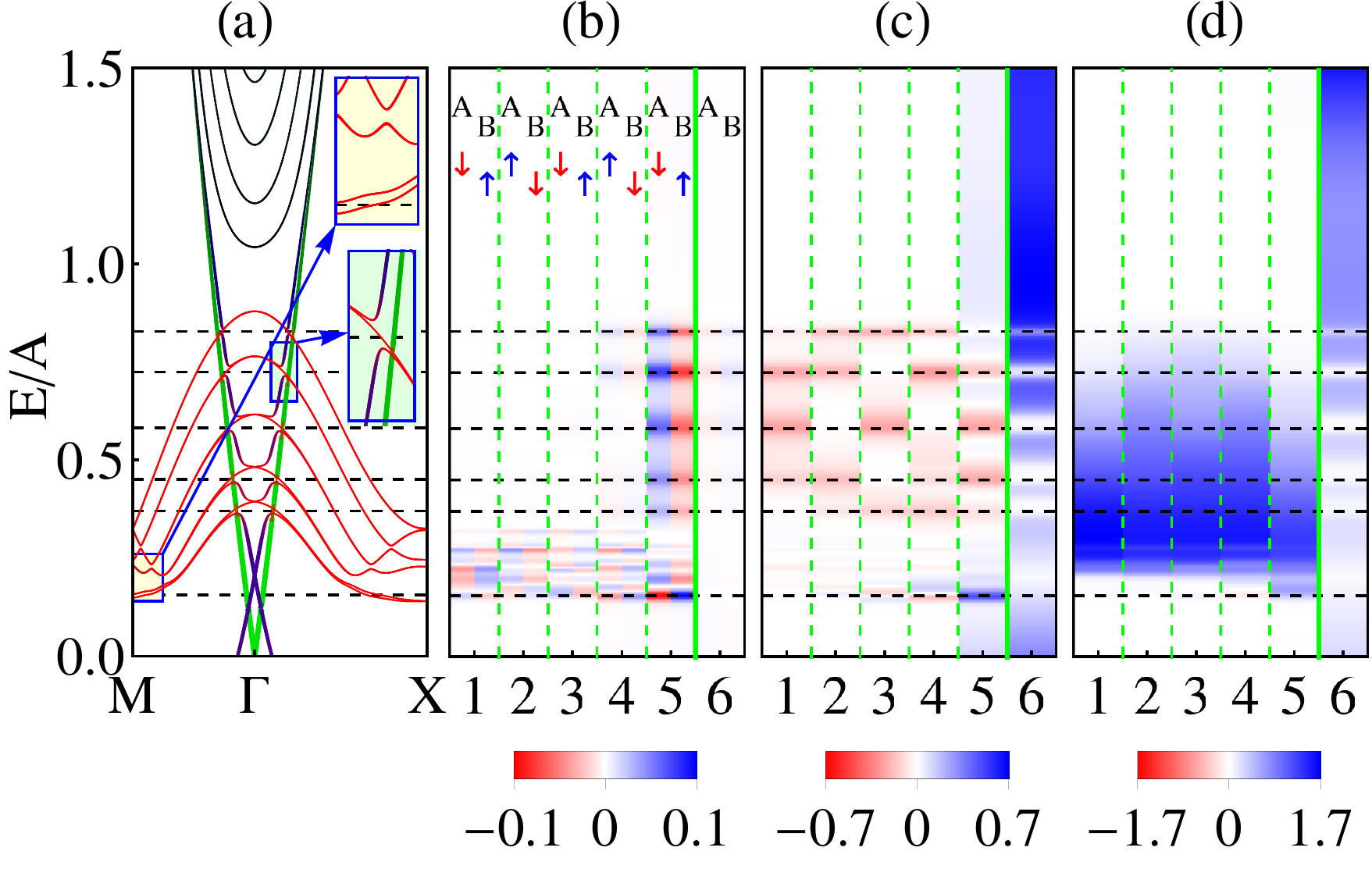}
\includegraphics[width=0.48\textwidth]{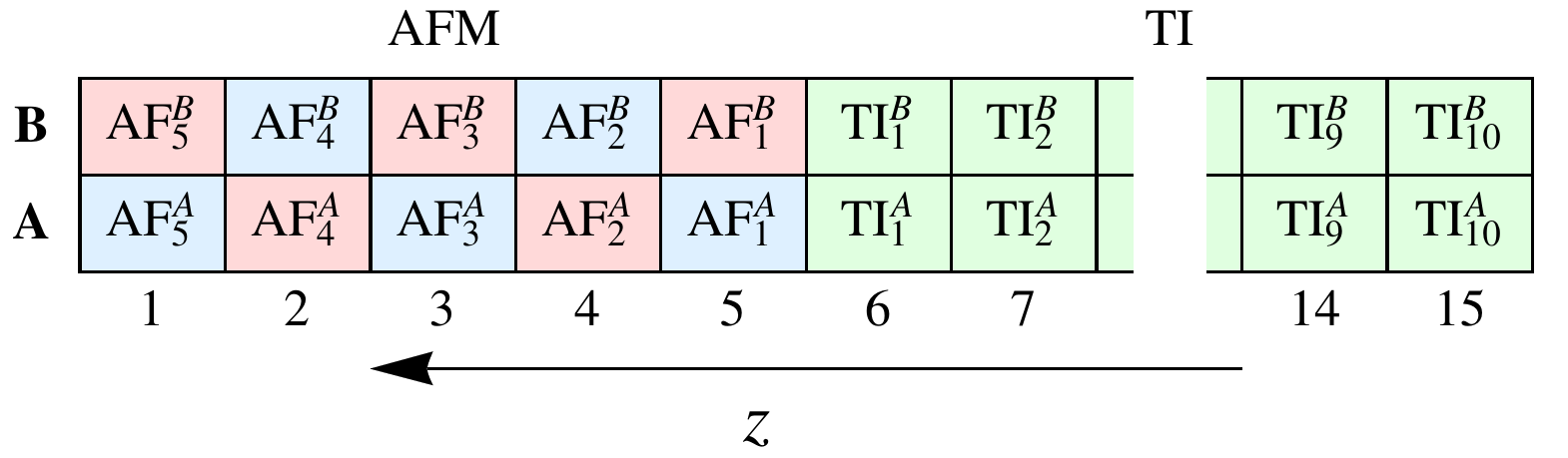}
\caption{(a) Band structure, non-equilibrium spin components (b) $S_x$ (c) $S_y$ and (d) conductivity for sites 1 to 12 with out-of-plane magnetization. The red, blue, green and black colors in (a) show the contribution from AF, interfacial TI, bottom TI and bulk TI bands.
The horizontal dashed lines denote the peaks of spin densities. In (b), (c) and (d) the vertical dashed green lines show the AF layers and the thick green line the position of the interface. Corresponding configuration is shown in the bottom panel where the blue and red boxes denote AF sites with up and down magnetic moments, and the green boxes correspond to TI sites.}
\label{fig:layer}
\end{figure}

First we calculate the site resolved spin density and conductivity for the AF layer and the interfacial TI layer (Fig.~\ref{fig:layer}). For simplicity we use natural units, i.e. $e=\hbar=a=1$ which does not affect the qualitative behavior of the observables. 

From Fig.~\ref{fig:band} we see that the coupling between TI and AF isolates one pair of AF bands from the conduction bands and brings it lower in the exchange gap region. This pair is dominated by the interfacial AF layer, which is also reflected in the fact that the spin densities and conductivity are stronger in the interfacial layer at this energy. Note that the spin densities show maxima at energies where the coupling between the AF and TI layers are strongest, which breaks the continuity of the TI bands (horizontal dashed lines in Fig.~\ref{fig:layer}a). Correspondingly, the spin densities in the TI layers reduce at these energies. Also note that at lower energy, the spin density has opposite sign. This is because the coupling moves the Dirac cone from top TI layer higher in energy and since the upper and lower halves of the Dirac cone possess opposite texture, the induced non-equilibrium spin density also switches sign.
To understand the connection between the equilibrium spin texture and non-equilibrium spin density, we calculate the site-resolved spin texture at different energies.

\begin{figure}[h]
\centering
\includegraphics[width=0.49\textwidth]{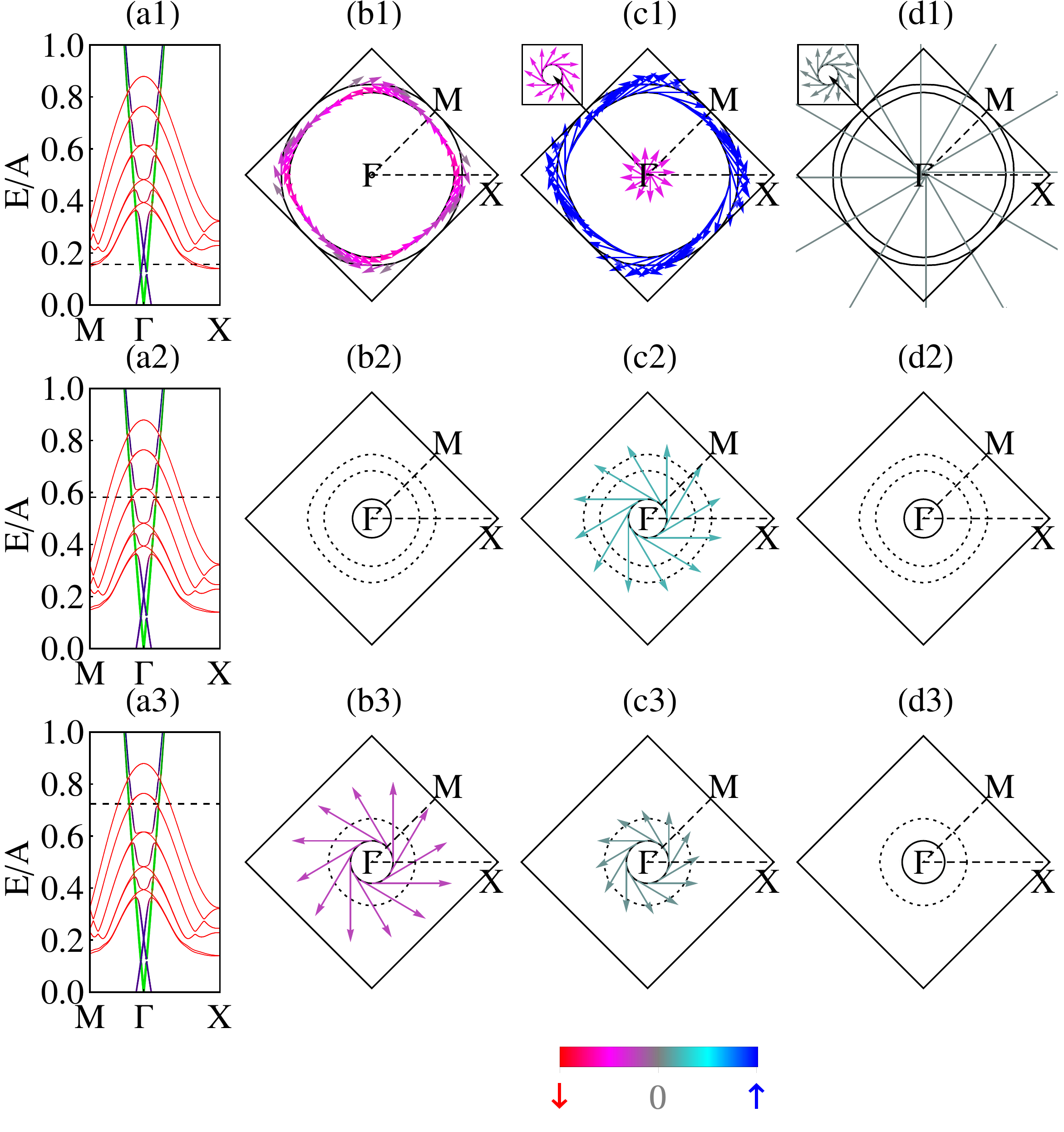}
\caption{(a) Band structure and spin texture at site (b) $\rm AF_2^B$, (c) $\rm AF_1^B$ and (d) $\rm TI_1^B$ (see bottom panel in Fig. \ref{fig:layer}) at different energies [(1) $E/A=0.156$, (2) $E/A=0.582$, (3)$E/A=0.724$]. The arrow indicates the in-plane component and the color of the arrow represents the out of plane component. The dotted lines correspond to the degenerate bands.}
\label{fig:texture}
\end{figure}

Figure \ref{fig:texture} shows the band structure and corresponding spin texture for site 8($\rm AF_2^B, \downarrow$), 10($\rm AF_1^B,\uparrow$) and 12($\rm TI_1^B$) for different energies. At $E/A = 0.156$, the degeneracy of the AF bands have been lifted due to coupling with TI layer and is manifested as two isolated rings in the Brillouin zone. Note that these two rings have opposite textures and in case of a degenerate band they cancel each other. This is why the non-equilibrium $S_y$, and therefore $S_x$, is zero in most of the regions. We further see that due to confinement effect, AF states at different energies are dominated by different layers, for example $E/A=0.582$ is dominated by the $10^{th}$ site ($5^{th}$ layer) whereas $E/A=0.724$ is dominated by the $8^{th}$ site ($4^{th}$ layer). This explains the variation of $S_y$ component at different layers (Fig.~\ref{fig:layer}). Note that at low energy the AF bands are well separated in momentum from the interfacial TI Dirac cone, which allows the TI layer to induce the SOC in AF layer and also to retain its own texture. 
From the slope of the band structure one can see that the TI and the AF layers have opposite group velocity within the intermediate energy range ($0.15<E/A<0.85$, which is roughly the region between the second and sixth horizontal dashed lines in Fig.~\ref{fig:layer}a). As a result, within this regime the AF layer and the TI layer possess opposite $S_y$ component in spite of having similar spin texture.

\subsection{Fermi surface and sea contributions to $S_x$}

\begin{figure}[h]
\centering
\includegraphics[width=0.48\textwidth]{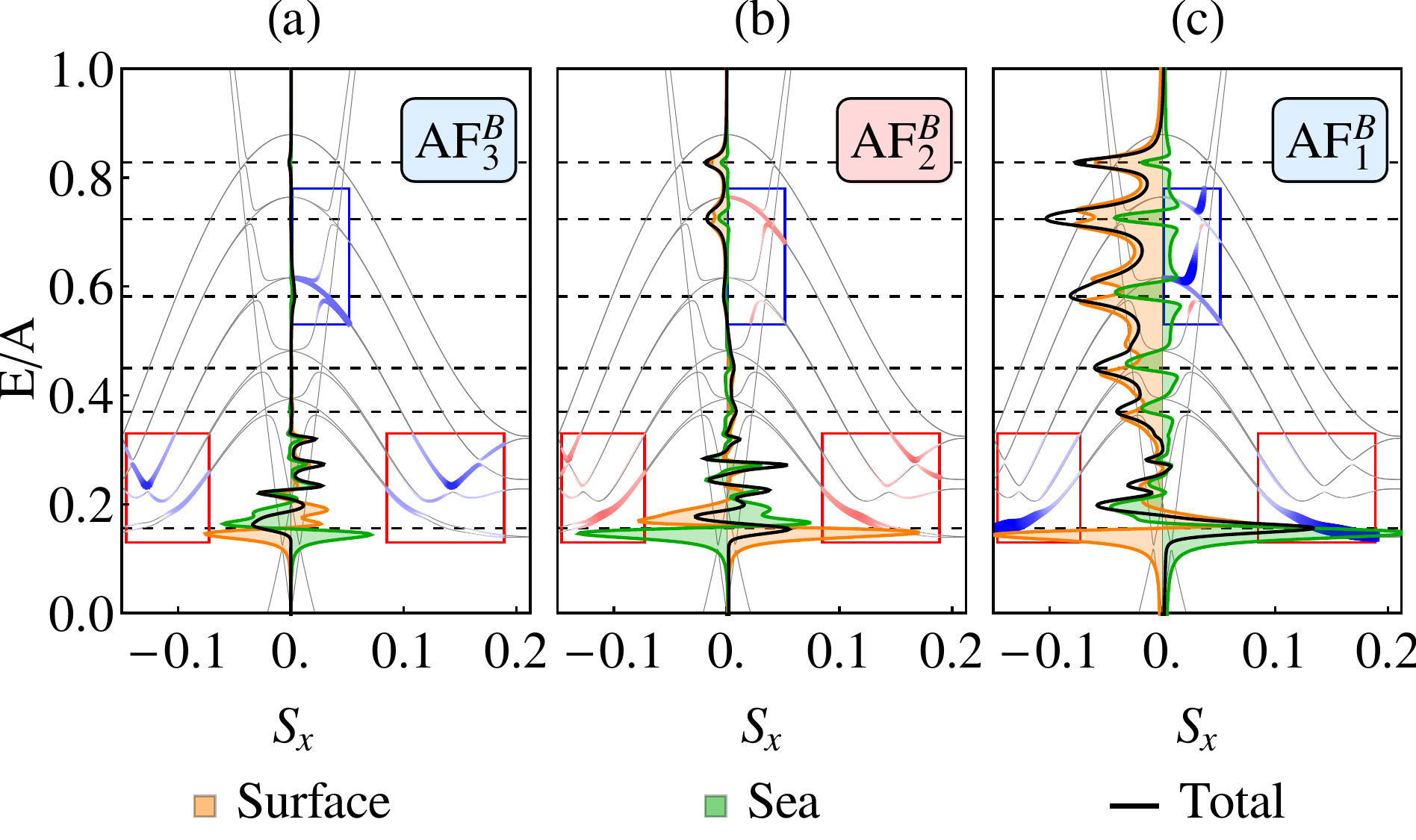}
\caption{Contribution from Fermi surface (orange) and Fermi sea (green) for the $B$ sublattice of first three AF layers. The equilibrium $S_z$ is shown in the boxed regions where red and blue color correspond negative and positive values. The maximum value in red box is 0.35 and maximum value in the blue box in 0.1.}
\label{fig:seasurface}
\end{figure}

The non-equilibrium $S_x$ component is particularly important because it enables the electrical control of the AF magnetic order. As a matter of fact, its magnitude depends both on the induced SOC as well as on the magnetization of the AF sublattices. As a result one can obtain a staggered $S_x$ that can be utilized to switch the antiferromagnetic order parameter \cite{Zelezny2014, Manchon2017}. Interestingly, unlike $S_y$ , $S_x$ has finite contribution from both Fermi surface [Eq. (\ref{w1})] and Fermi sea [Eq. (\ref{w2})] (see Fig.~\ref{fig:seasurface}) which indicates its topological origin. 

One can readily see that both of the sea and surface terms dominate in the region where the degeneracy of the AF band is lifted. Note that the Fermi sea contribution attains a maximum value in the middle of the gap where the surface contribution shows a minima. 
This is because while the Fermi sea term depends on the strength of Berry curvature, the Fermi surface term depends on the induced SOC. To understand the topological origin of the non-equilibrium $S_x$ one needs to study the relevant topological invariant. However,  constructing a proper topological invariant for the interfacial states of such antiferromagnetic heterostructure is mathematically quite challenging and is beyond the scope of present work. Instead, here we present a heuristic argument to determine the  Berry curvature contribution. We note that the Berry curvature is analogous to a magnetic field whose strength is maximum when the spin texture encounters a singularity marked by a zero in-plane and finite out-of-plane component. To demonstrate that we consider the Rashba-Ferromagnet Hamiltonian
\begin{eqnarray}
H_R=\hbar^2 k^2/2m_0+\alpha \hat{z}.(\bm{\sigma} \times \bm{p}) + \Delta \sigma_z.
\label{RFM}
\end{eqnarray}

\begin{figure}[h]
\centering
\includegraphics[width=0.43\textwidth]{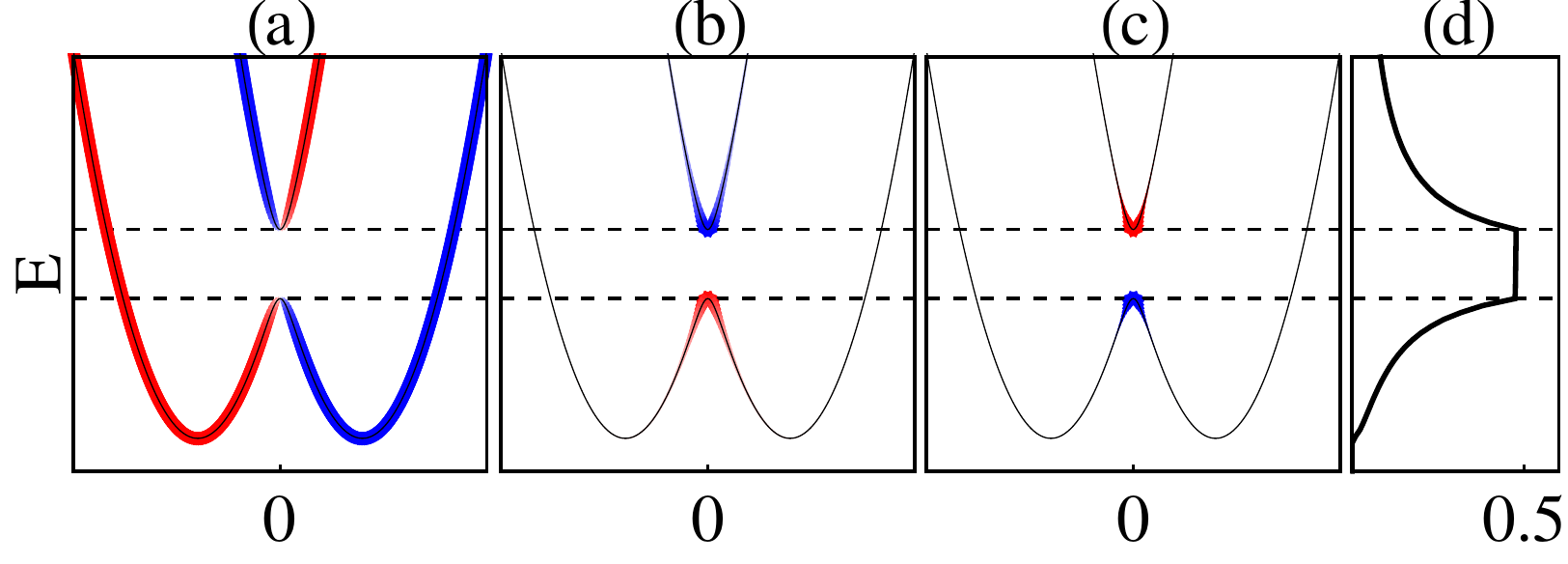}
\caption{Band structure of a Rashba-ferromagnet system (\ref{RFM}) along $k_x$ with projected (a) $\sigma_y$, (b) $\sigma_z$, (c) Berry curvature and (d) integrated Berry curvature or the anomalous Hall coefficient. The dashed lines show the exchange gap. Here we use $\hbar=1$, $m_0=0.5$, $\alpha=1$ and $\Delta=0.05$.}
\label{fig:berry}
\end{figure}

Following Ref. \onlinecite{Culcer2003} the Berry curvature reads
\begin{eqnarray}
\Omega^\pm(k)= \mp \frac{1}{2} \frac{\alpha^2 \Delta}{ (\Delta^2+\alpha^2k^2)^{3/2}},
\label{RBerry}
\end{eqnarray}
which can readily be compared with the momentum-dependent z-component of the spin density,
\begin{eqnarray}
S^\pm_{z}(k)= \pm \frac{\Delta}{(\Delta^2+\alpha^2k^2)^{1/2}},
\label{RSz}
\end{eqnarray}
where $\pm$ corresponds two different bands. It clearly appears that the maximum in Berry curvature coincides with a maximum in $S_z$. Corresponding band structure along with its spin projection and Berry curvature is shown in Fig.~\ref{fig:berry}. One can readily see that the Berry curvature is localized in a region where the out-of-plane component of spin is maximum, consistently with Eq. (\ref{RBerry},\ref{RSz}) above. At these points the in-plane texture vanishes creating a singularity. Therefore we track the equilibrium $S_z$ as a smoke signal for Berry curvature.  
From Fig.~\ref{fig:seasurface}, we see that most of the equilibrium $S_z$ is localized at the bottom of the positive AF bands and therefore most of the non-equilibrium $S_x$ component is also generated in this region. Since the bands with opposite texture have same $S_z$ component (Fig.~\ref{fig:texture}c1), the Fermi surface component switches sign as we move from the first AF band to the second ($E/A\sim0.156$). Corresponding Berry curvature is localized at the bottom of the bands where the Fermi sea term changes sign. At higher energy ($E/A\sim0.724$) opposite textures possess opposite $S_z$ (see the blue box in Fig.~\ref{fig:seasurface}) and therefore the Fermi surface term does not change sign anymore. The Fermi sea term being dependent on the Berry curvature still oscillates with the sign of $S_z$. For a decoupled AF band, there is no equilibrium $S_z$ component near the band maxima. In an AF-TI heterostructure this component, and therefore its corresponding Berry curvature is produced due to the interaction between the TI and AF and therefore is localized close to the interface. As a result, for $E/A>0.3$, the non-equilibrium $S_x$ component is finite only near the interface (Fig.~\ref{fig:layer}b) although the $S_y$ component is visible deep inside (Fig.~\ref{fig:layer}c). This is also reflected in the fact that the Fermi sea term dominates in the bottom of the positive AF band for the  first AF layer ($\rm AF_1^B$, Fig.~\ref{fig:seasurface}c), whereas for $\rm AF_2^B$ onwards, the surface term dominates. As a result, while in $\rm AF_1^B$ the $S_x$ component has the same sign as the local magnetic moment, $\rm AF_2^B$ and $\rm AF_3^B$ have opposite signs.

\subsection{Effect of AF-TI coupling}

To understand the cumulative behavior we introduce the staggered x-component of the spin density, $S_x^{stg}$, the total y-component, $S_y^{tot}$, and the total conductivity, $\sigma_{xx}^{tot}$, defined as
\begin{eqnarray}
S_x^{stg} &=& \sum_i^\uparrow S_x^i - \sum_j^\downarrow S_x^j \\
S_y^{tot} &=& \sum_i^\uparrow S_y^i + \sum_j^\downarrow S_y^j\\
\sigma_{xx}^{tot} &=& \sum_i^\uparrow \sigma_{xx}^j + \sum_j^\downarrow \sigma_{xx}^i 
\end{eqnarray} 
where the indices $i$ and $j$ run over the AF sites with up and down spin. 
The $S_x^{stg}$ component induces the damping like torque that enables the electrical control of the Ne\'el order parameter \cite{Gomonay2010, Zelezny2014, Manchon2017}, while the $S_y^{tot}$ induces an effective field that does not contribute to the switching.
 First we look at the effect of coupling between the AF and TI which is the only way to imprint SOC on the AF layer and produce the non-equilibrium spin density.

\begin{figure}[h]
\centering
\includegraphics[width=0.49\textwidth]{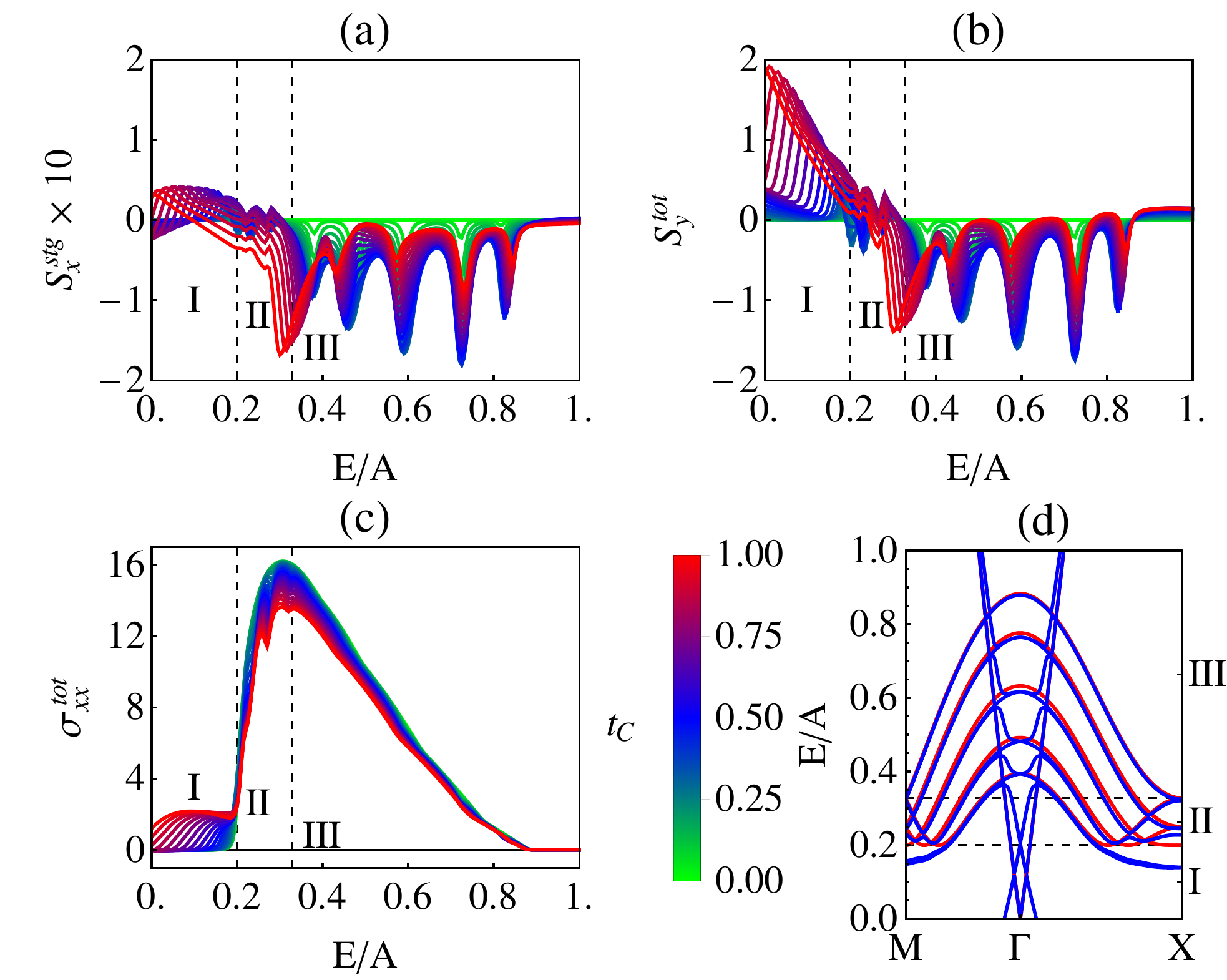}
\caption{(a) $S_x^{stg}$, (b) $S_y^{tot}$ and (c) $\sigma_{xx}^{tot}$ of the AF layer as a function of coupling strength whose value is shown in the colorbar. (d) The band structure of decoupled (red) AF-TI and with coupling $t_C=0.5A$ (blue). The dashed lines show three different regions I, II and III.}
\label{fig:coupling}
\end{figure}

Figure \ref{fig:coupling} shows the variation of  $S_x^{stg}$, $S_y^{tot}$ and $\sigma_{xx}^{tot}$ as a function of energy and with respect to the coupling strength between AF and TI layers ($t_C$). For better understanding we divide the energy range into three  regions. Region I corresponds to the exchange gap of the decoupled AF bands. This region can be occupied by an AF band only upon turning on the coupling between AF and TI (Fig.~\ref{fig:band}). Region II spans the energy range corresponding to the bottom of the positive AF bands which corresponds to the maximum density of states and also contains strong equilibrium $S_z$ component. The coupling opens a gap between different bands in this region (Fig.~\ref{fig:band}). Region III  contains all the energies above region II. The coupling mainly distorts the TI bands rather than the AF bands here. $S_y^{tot}$ increases in both region I and region II as it depends only on the induced SOC. Note that the $S_x$ component switches sign with respect to the local magnetization in the first two consecutive layers (Fig.~\ref{fig:layer}), which reduces $S_x^{stg}$. This effect is more prominent in region I as it is dominated both by the first ($\rm AF_1^{A,B}$) and second ($\rm AF_2^{A,B}$) interfacial as well as bulk AF layer (Fig.~\ref{fig:layer}). This is also reflected in the equilibrium spin texture (Fig.~\ref{fig:texture}). As a result, in region I,  $S_x^{stg}$ starts decreasing for stronger coupling strength (Fig.~\ref{fig:coupling}a). Region II is less dominated by the interfacial AF layer and therefore the cancellation due to opposite texture is minimized. Moreover stronger coupling opens larger gaps which provides more $S_z$ component (i.e. stronger Berry curvature) giving rise to larger value of $S_x^{stg}$. In region III the coupling breaks the degeneracy of the AF bands which ensures that only one type of texture dominates. For smaller coupling two opposite textures have almost same contribution and cancel each other. For stronger coupling the TI bands undergo more distortion resulting in reduced SOC and weak spin texture. Therefore we see that in region III both $S_y^{tot}$ and $S_x^{stg}$ attain a maximum value for some intermediate coupling strength. The conductivity on the other hand depends on the density of states and increases in region I and decreases in region II and III with coupling strength as the coupling pulls down the AF states from region II and III to region I (Fig.~\ref{fig:coupling}).

\subsection{Coupling to surface vs bulk TI states}

So far we have considered the AF bands coupled to the surface TI bands only. In reality the AF bands can be connected to the bulk TI bands as well or can be coupled to both surface and bulk. To understand these different scenarios we change the position of the AF bands by varying the onsite energy [$\varepsilon_0$, Eq. (\ref{H_AF})] of the AF layer.

\begin{figure}[h]
\centering
\includegraphics[width=0.49\textwidth]{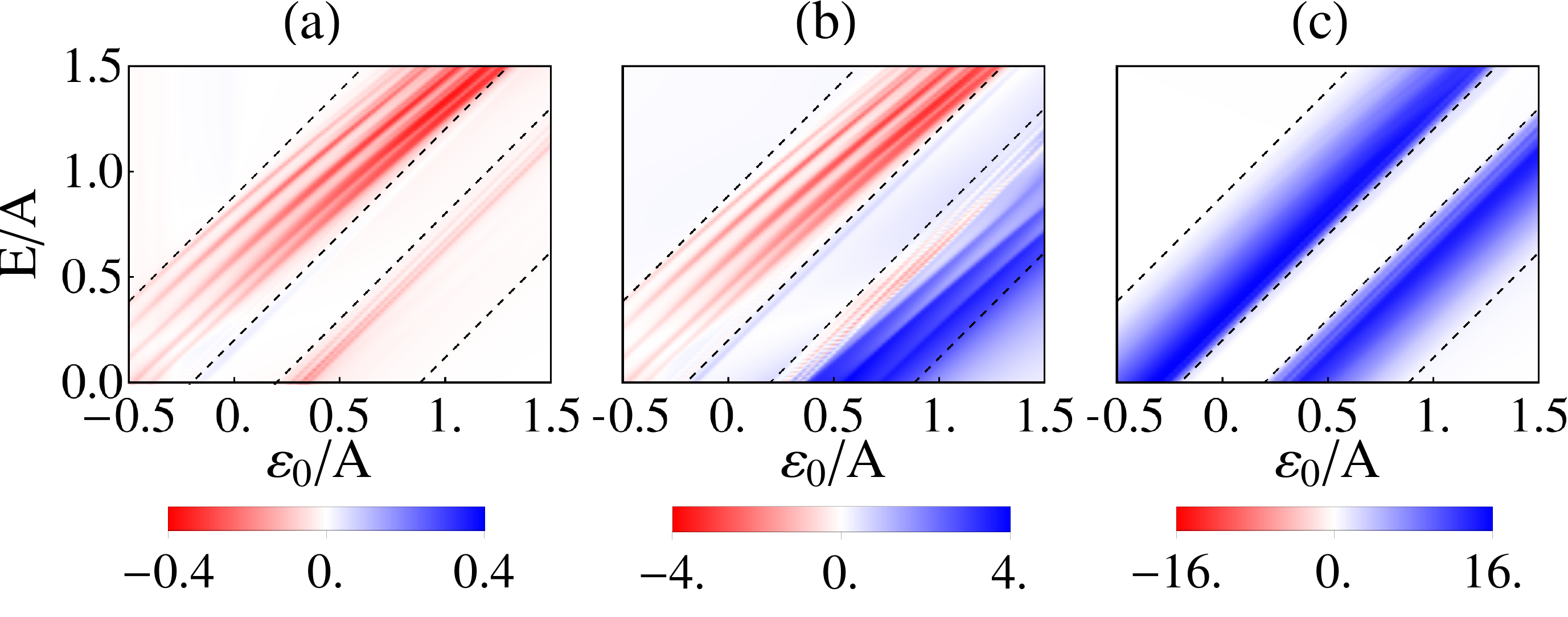}
\caption{Energy dependence of (a) $S_x^{stg}$, (b)  $S_y^{tot}$ and (c) $\sigma_{xx}^{tot}$ of the AF layer as a function of the onsite energy of the AF layer with $t_C=0.5A$. The inclined dashed line shows the region spanned by the decoupled AF bands.}
\label{fig:e0}
\end{figure}

Fig.~\ref{fig:e0} shows the variation of $S_x^{stg}$,  $S_y^{tot}$ and $\sigma_{xx}^{tot}$ for different onsite energies $\varepsilon_0$. The total conductivity ($\sigma_{xx}^{tot}$) is confined within the region marked by the dashed line denoting the energy range occupied by the AF bands. For $S_x^{stg}$ and $S_y^{tot}$, one can clearly see five branches corresponding to the intersections of AF bands with the TI bands at lower energies. Careful observation reveals a faint branch with opposite sign within the exchange gap region which corresponds to the interfacial AF band. For larger values of $\varepsilon_0$ the AF bands are coupled to more number of bulk TI bands which increases both $S_y$ and $S_x$. However one should note that in this case more current flows through the bulk TI, which effectively reduces the spin Hall angle and hence the efficiency of the system \cite{Ghosh2018}. Besides, the bulk bands are not immune to the scattering which makes the coupling to the bulk bands inefficient against impurity scattering.

\subsection{Effect of impurity}

\begin{figure}[h]
\centering
\includegraphics[width=0.49\textwidth]{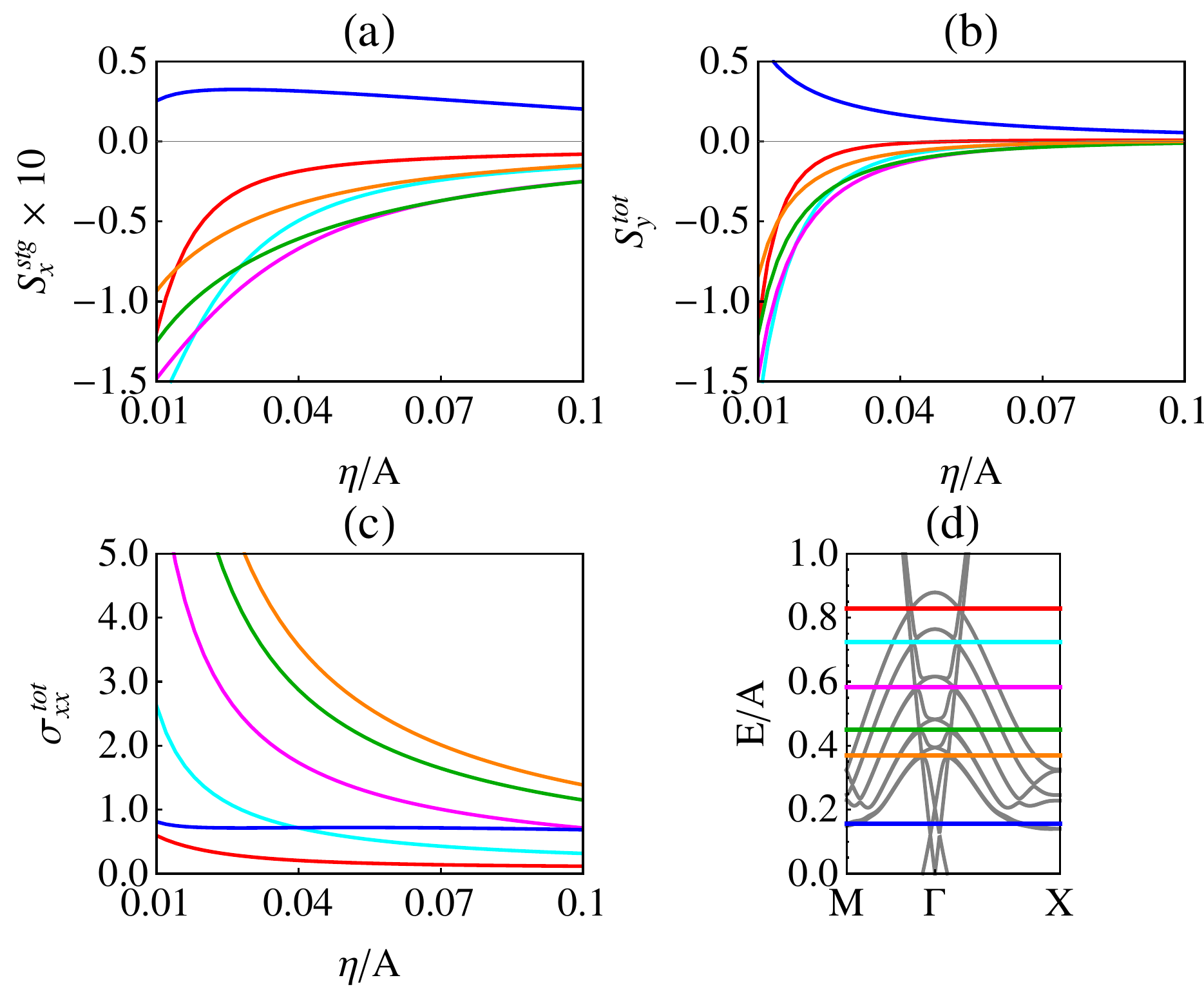}
\caption{(a) $S_x^{stg}$, (b) $S_y^{tot}$ and (c) $\sigma_{xx}^{tot}$ as a function of broadening. Different line colors correspond to different energies as shown in panel (d) along with the band structure.}
\label{fig:broadening}
\end{figure}

Next we study the effect of impurity scattering modelled as a constant broadening, an approach that is found to be sufficient in realistic systems as well \cite{Freimuth2014}. Figure \ref{fig:broadening} shows the variation of $S_x^{stg}$, $S_y^{tot}$ and $\sigma_{xx}^{tot}$ with the broadening parameter ($\eta$). We have already seen that the non-equilibrium $S_x$ and $S_y$ are maximum only at some specific values of energies (Figs. \ref{fig:layer},\ref{fig:coupling}). Here we consider the energies marked by the dashed line in Fig.~\ref{fig:layer} and Fig.~\ref{fig:seasurface}.

From Fig.~\ref{fig:broadening} we see that for energies coming from region III (Fig.~\ref{fig:coupling}), $S_x^{stg}$, $S_y^{tot}$ and $\sigma_{xx}^{tot}$ show a $1/\eta$ decay which is the characteristic of a metallic system. However the behavior is completely different when we choose an energy from the region I (see Fig.~\ref{fig:broadening}, blue line). We see that $\sigma_{xx}^{tot}$ remains constant as the effect of impurity is quenched by the strong spin polarization. $S_y^{tot}$ also falls down but with a much slower rate which indicates the weakening of the induced SOC. $S_x^{stg}$ shows an initial increase and then becomes almost constant which also points towards its topologically protected origin. Note that the enhancement of $S_x^{stg}$ component due to scattering has also been observed in two dimensional antiferromagnetic TIs \cite{Ghosh2017} as well as TI-ferromagnet heterostructures \cite{Ghosh2018}.

\subsection{Angular dependence of torques}

\begin{figure}[h]
\centering
\includegraphics[width=0.49\textwidth]{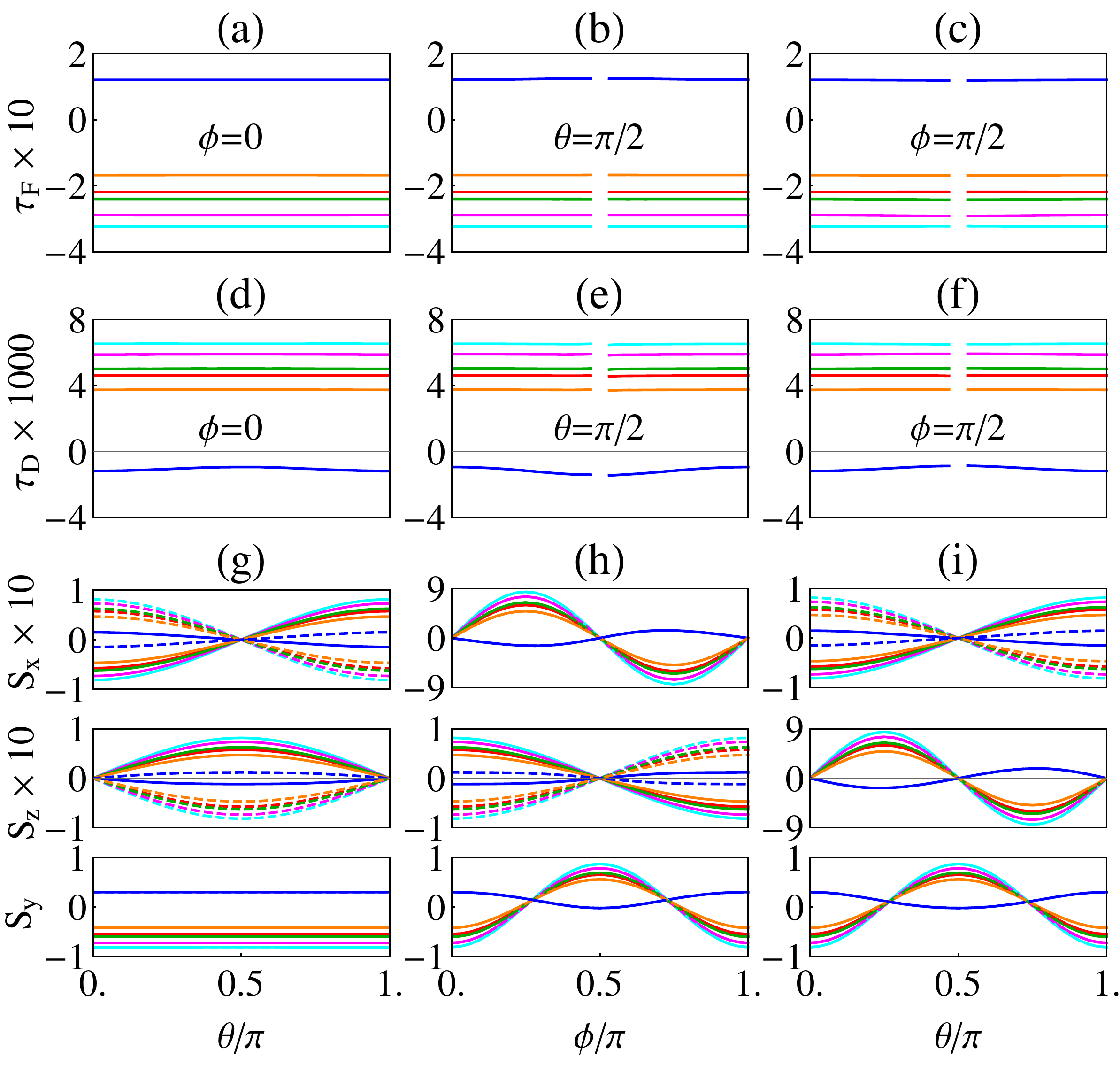}
\caption{Variation of $\bar\tau_{FL}$(a,b,c) and $\bar\tau_{DL}$(d,e,f) with respect to $\theta$ (a,c;d,f) and $\phi$ (b;e). The inset value shows the fixed angle. (g), (h) and (i) shows $S_x$, $S_z$ and $S_y$ where solid and dashed line denotes the contribution from AF sites with up and down spin. The colors represent different energies as described in Fig.~\ref{fig:broadening}d.}
\label{fig:angle}
\end{figure}

Finally we present the angular dependence of field like and damping like torques with respect to the polar ($\theta$) and azimuthal ($\phi$) angles (Fig.~\ref{fig:angle}) for different planes. Note that in the case of out-of-plane magnetization the field like and damping like torques are directly proportional to $S_y^{tot}$ and $S_x^{stg}$. Therefore one can easily understand the nature of field like and damping like torques from the  $S_x^{stg}$ and $S_y^{tot}$ components. In general the torque components at each site  can be expressed as 
\begin{eqnarray}
\bm{T}^i &=& \bm{m}^i \times \bm{S}^i \nonumber \\
&=& \tau^i_{F} (\bm{m}^i \times \bm{e_y}) + \tau^i_{D} \bm{m}^i \times (\bm{m}^i \times \bm{e_y})\nonumber \\
\end{eqnarray}
$\bm{e_x}$ is direction of current flow and $\bm{e_z}$ is normal to the surface. The general expressions for $\tau^i_{F,D}$ for any arbitrary angle are given by


\begin{eqnarray}
\tau^i_{F} &=& (\bm{m}^i \times \bm{S}^i) \cdot (\bm{m}^i \times \bm{e_y}) /|\bm{m}^i \times \bm{e_y}|^2 \nonumber \\
&=& -\frac{\sin \phi  \cos \phi  \sin^2 \theta}{1-\sin^2 \phi \sin^2 \theta} S^i_x + S^i_y - \frac{\sin\phi \sin \theta \cos\theta}{1-\sin^2 \phi \sin^2 \theta} S^i_z, \nonumber \\ \label{tf} \\
\tau^i_{D} &=& (\bm{m}^i \times \bm{S}^i) \cdot (\bm{m}^i \times (\bm{m}^i \times \bm{e_y})) /|\bm{m}^i \times (\bm{m}^i \times \bm{e_y})|^2 \nonumber \\
&=& - \frac{\cos \theta}{1-\sin^2 \phi \sin^2 \theta} S^i_x + \frac{\sin \theta \cos \phi}{1-\sin^2 \phi \sin^2 \theta} S^i_z. \label{td}
\end{eqnarray}
\begin{table}[h]
\centering
\begin{tabular}{|l|c|c|}
\hline
Plane & $\tau^i_{F}$ & $\tau^i_{D}$ \\ \hline 
$zx (\phi=0)$ & $S^i_y$ & -$S^i_x \cos \theta + S^i_z \sin \theta $ \\ \hline
$xy (\theta=\pi/2)$ & $S^i_y - S^i_x \tan \phi $ & $S^i_z /\cos \phi $ \\ \hline
$xy (\phi=\pi/2)$ & $S^i_y - S^i_z \tan \theta $ & $S^i_x /\cos \theta $ \\ \hline
\end{tabular}
\caption{Torque coefficients $\tau^i_{F}$ and $\tau^i_{D}$ for $zx$, $xy$ and $yz$ plane.}
\label{nfd}
\end{table}

Note that each layer contains two sites with opposite magnetization. Since these two sites have same sign of induced SOC, they would have same sign of field like coefficient ($\tau_F$) and opposite sign of damping like coefficient ($\tau_D$). Therefore we define the total field like and damping like torque coefficients as
\begin{eqnarray}
\tau_F &=& \sum_i^\uparrow \tau_F^i + \sum_j^\downarrow \tau_F^j \\
\tau_D &=& \sum_i^\uparrow \tau_D^i - \sum_j^\downarrow \tau_D^j. 
\end{eqnarray}

\begin{figure}[h]
\centering
\includegraphics[width=0.49\textwidth]{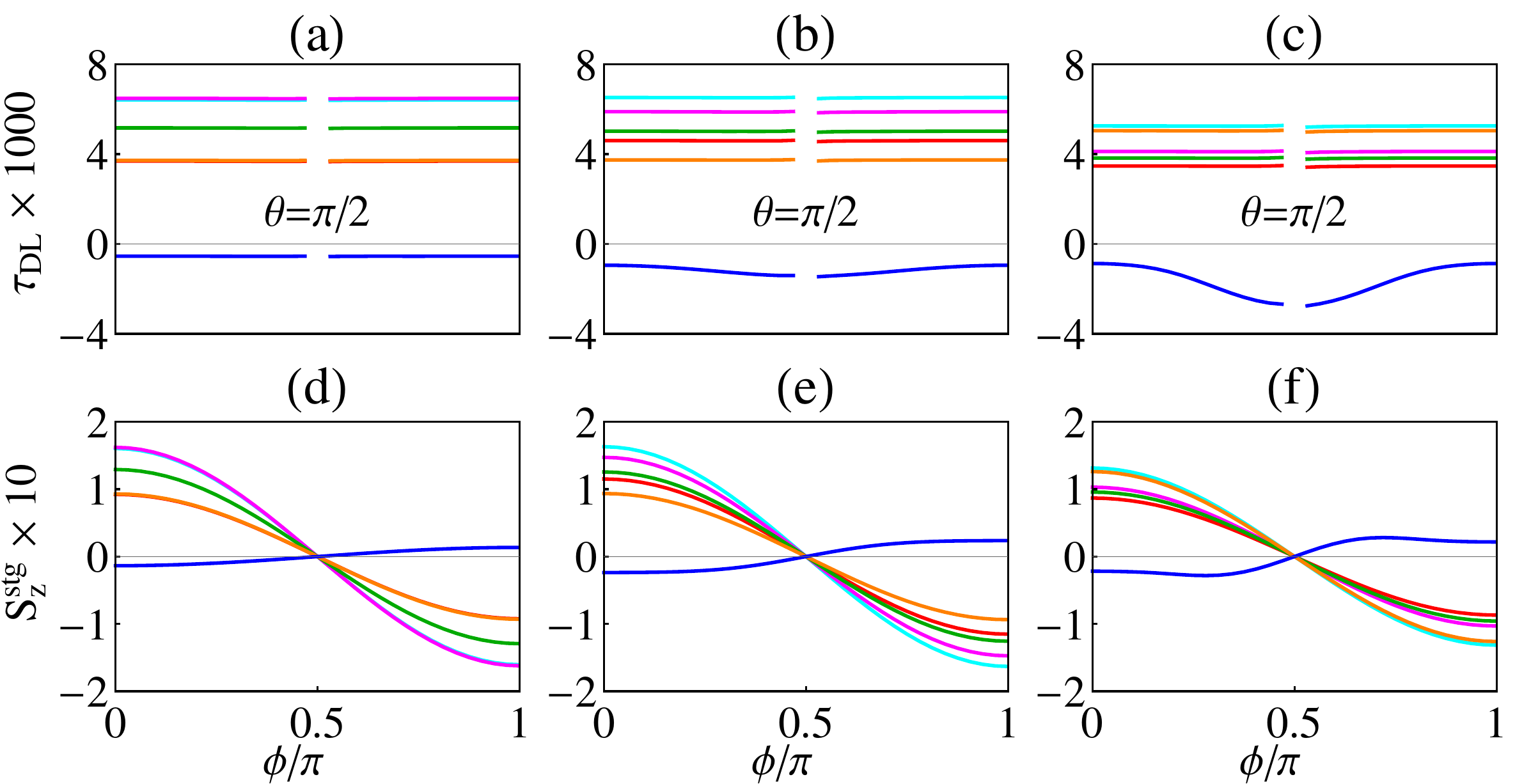}
\caption{Variation of $\tau_{D}$ in $xy$ plane for (a) $t_C=0.25$, (b)  $t_C=0.50$  and (c)  $t_C=0.75$ and corresponding non-equilibrium $S_z$ (d,e,f). Different colors correspond the energies at which $S_z$ have peaks. The band structures are shown in Fig. \ref{fig:band}.}
\label{fig:dz}
\end{figure}

From Fig. \ref{fig:angle} one can readily see that  both $\tau_F$ and $\tau_D$ do not show any angular dependence for all energies dominated by the bulk AF layers (region II and III in Fig. \ref{fig:coupling}). 
When the magnetization is in the $xz$ plane (Fig.~\ref{fig:angle}a,d), it does not affect the non-equilibrium $S_y$ component. As a result $\tau_{F}$ remains constant for all energies. Since $S_{x,z}$ is generated through the interaction with $z,x$ components of magnetization, they change as $\cos\theta,\sin\theta$ respectively and therefore $\tau_{D}$ remains constant. The situation is different for $xy$ and $yz$ planes.
In these cases, the non-equilibrium $S_y$ component also has a contribution from the $y$ component of magnetization. In addition the texture itself is modified resulting in a complex angular dependence. Note that for $xy,yz$ planes, the periodicity of $S_{y}$ and $S_{x,z}$ seems to be double compared to the periodicity of $S_{z,x}$. Careful observation shows that the variation is not strictly sinusoidal and depends on the strength of the induced SOC. As a result the blue lines undergo less distortion due to their stronger SOC compared to the other lines.  
The $S_{z,x}$ component, on the other hand, is generated through the interaction with $x,z$ component of magnetization and therefore shows a $\cos \phi,\theta$ dependence. Note that in case of $\tau_D$, there is small deviation from constant value for the blue line in all three cases. This effect comes from the \textit{``breathing''} of the band structure due to the interplay between magnetism and SOC \cite{Zelezny2017}.
 For $\tau_F$ this effect is nullified by the contribution from $S_y$ and $S_{x,z}$. This is further clarified by showing the angular dependence of $\tau_D$ and $S_z$ in $xy$ plane for different coupling strengths (Fig. \ref{fig:dz}).

\section{Conclusion}
In this article we present a systematic study of non-equilibrium spin density and SOT for an AF-TI heterostructure using a tight binding model within the framework of linear response theory. From the site resolved spin density we show the existence of a staggered $S_x$ component which can be utilized to switch the antiferromagnetic order parameter. We explain the behavior of different spin components at different energies and show their connections with the band structure and equilibrium spin texture. We show that for an out-of-plane magnetization, the $S_x$ component is dominated by both Fermi sea and Fermi surface contributions. The contribution from the Fermi sea points towards its topological origin, which is later verified from the impact of impurity. We further study the effect of the coupling between the AF and TI layers and show that the $S_x$ component attains its maximum value for an intermediate coupling strength whereas $S_y$ component shows different behavior at different energies. The non-equilibrium spin density also depends on the total amount of overlap which we demonstrate by coupling the AF bands with both the surface and the bulk TI bands. However the spin density produced away from the Dirac cone of the TI is not robust against impurity and falls down rapidly with growing impurity strength. Interestingly we find that near the Dirac cone the $S_x$ component is slightly enhanced by the impurity. At this energy, the conductivity is not affected by the impurity. Finally we show the angular dependence of the non-equilibrium spin densities as well as the different torque coefficients, which have the same behavior as a two dimensional Rashba gas for the bulk AF bands. However within the AF gap, the damping like torque shows a complex angular dependence due to the breathing of band structure which can be enhanced with the coupling strength between AF and TI. 
 
\section*{Acknowledgement}
The authors would like to acknowledge the computation resources in supercomputer Shaheen. This work is supported by the King Abdullah University of Science and Technology. 

\bibliographystyle{apsrev4-1}
\bibliography{AFTI1}
\end{document}